
\input harvmac
\overfullrule=0pt

\def\Title#1#2{\rightline{#1}
 \ifx\answ\bigans \nopagenumbers\pageno0\vskip1in 
  \baselineskip 15pt plus 1pt minus 1pt
 \else\def\listrefs{\footatend\vskip 1in\immediate\closeout\rfile\writestoppt
  \baselineskip=14pt\centerline{{\bf References}}\bigskip{\frenchspacing%
  \parindent=20pt\escapechar=`\input refs.tmp\vfill\eject}\nonfrenchspacing}
  \pageno1\vskip.8in
 \fi \centerline{\titlefont #2}\vskip .5in}

\newcount\figno \figno=0
\input epsf
\def\fig#1#2#3{\par\begingroup\parindent=0pt\leftskip=1cm\rightskip=1cm
 \baselineskip=11pt \global\advance\figno by 1 \midinsert
 \epsfxsize=#3 \centerline{\epsfbox{#2}} \vskip 12pt
 {\bf Fig. \the\figno:} #1\par \endinsert\endgroup\par }
\font\blackboard=msbm10 
\font\blackboards=msbm7 \font\blackboardss=msbm5
\newfam\black \textfont\black=\blackboard
\scriptfont\black=\blackboards \scriptscriptfont\black=\blackboardss
\def\Bbb#1{{\fam\black\relax#1}}
\def\bC{{\Bbb C}}
\def\bZ{{\Bbb Z}}
\def\hepth#1{{hep-th/#1}}

\def\PRD#1#2#3{{\sl Phys.\ Rev.}\ {\bf D#1}(#2)#3}
\def\NPB#1#2#3{{\sl Nucl.\ Phys.}\ {\bf B#1}(#2)#3}
\def\PLB#1#2#3{{\sl Phys.\ Lett.}\ {\bf #1B}(#2)#3}
\def\PRL#1#2#3{{\sl Phys.\ Rev.\ Lett.}\ {\bf #1}(#2)#3}

\Title{CLNS-97/1492}
{\vbox{\centerline{S-Duality and Global Symmetries} \medskip
\centerline{in N=2 Supersymmetric Field Theory}}}
\bigskip
\centerline{Philip C. Argyres}
\smallskip
\centerline{\it Newman Laboratory,
Cornell University, Ithaca NY 14853-5001 USA}
\centerline{\tt argyres@mail.lns.cornell.edu}
\bigskip
\baselineskip=14pt plus 2pt minus 1pt
\noindent
S-dualities in scale invariant $N=2$ supersymmetric field theories
are derived by embedding those theories in asymptotically free $N=2$
theories with higher rank gauge groups.  S-duality transformations on
the coupling of the scale invariant theory follow from global
symmetries acting on the Coulomb branch of the higher rank theory.
Since these global symmetries are exact in the asymptotically free
theory, this shows that S-duality is an exact equivalence of
$N=2$ theories and not just a property of their supersymmetric
states.
\Date{6/97}

\nref\MO{C. Montonen and D. Olive, \PLB{72}{1977}{117}.  E. Witten and 
	D. Olive, \PLB{78}{1978}{97}.  H. Osborn, \PLB{83}{1979}{321}.}
\nref\Se{A. Sen, \hepth{9402032}, \PLB{329}{1994}{217}.  S. Sethi, M. Stern 
	and E. Zaslow, \hepth{9508117}, \NPB{457}{1995}{484}.  J. Gauntlett 
	and J. Harvey, \hepth{9508156}, \NPB{463}{1996}{287}.  F. Ferrari, 
	\hepth{9702166}.  K. Lee and P. Yi, \hepth{9706023}.}
\nref\VW{C. Vafa and E. Witten, \hepth{9408074}, \NPB{432}{1994}{3}.}
\nref\SWii{N. Seiberg and E. Witten, \hepth{9408099}, \NPB{431}{1994}{484}.} 
\nref\APS{P. Argyres, M. Plesser and A. Shapere, \hepth{9505100}, 
	\PRL{75}{1995}{1699}.}  
\nref\AS{P. Argyres and A. Shapere, \hepth{9509175}, \NPB{461}{1996}{437}.}
\nref\ASYT{O. Aharony, C. Sonnenschein, S. Yankielowicz and S. Theisen,
	\hepth{9611222}, \NPB{297}{1997}{177}.  M. Douglas, D. Lowe and 
	J. Schwarz, \hepth{9612062}, \PLB{394}{1997}{297}.}
\nref\HMS{J. Harvey, G. Moore and A. Strominger, \hepth{9501022}, 
	\PRD{52}{1995}{7161}.  M. Bershadsky, A. Johansen, V. Sadov and 
	C. Vafa, \hepth{9501096}, \NPB{448}{1995}{166}.} 
\nref\WiMthNii{E. Witten, \hepth{9703166}.  K. Lansteiner, E. Lopez and 
	D. Lowe, \hepth{9705199}.  A. Brandhuber, J. Sonnenschein, S. 
	Theisen and S. Yankielowicz, \hepth{9705232}.}
\nref\JO{I. Jack and H. Osborn, \NPB{343}{1990}{647}.  H. Sonoda, 
	\hepth{9205085}, \NPB{383}{1992}{173}; \hepth{9205084}, 
	\NPB{394}{1993}{302}.  B. Dolan, \hepth{9403070}, {\sl Int.\ J.\ 
	Mod.\ Phys.}\ {\bf A10}(1995)2439.}
\nref\ArPol{P. Argyres, \hepth{9705076}.}
\nref\SWi{N. Seiberg and E. Witten, \hepth{9407087}, \NPB{462}{1994}{19}.} 

S-duality refers to a quantum equivalence of classically inequivalent
field theories.  This equivalence often consists of the identification
of a theory with the same theory deformed by an exactly marginal
operator.  These deformations can be viewed as transformations on the
classical space of couplings of the theory, and the duality group is
the group of such transformations.  Since the elements of the duality
group are supposed to connect equivalent theories, the quantum
coupling space is the classical one divided by the action of the
duality group.  The paradigmatic example \MO\ is the strong-weak
coupling duality of $N=4$ supersymmetric Yang-Mills theory under which
theories with couplings $\tau$ and $-1/\tau$ are identified.  Since
the $N=4$ theory is scale invariant the exactly marginal operator
corresponds to changing the coupling.

This letter discusses related S-dualities for scale invariant $N=2$
supersymmetric field theories.  Up to now, the evidence for
S-dualities in field theory has come from the spectrum of BPS
saturated states \Se\ and from low-energy effective actions
\refs{\VW-\ASYT}.  This evidence leaves open the possibility that the
non-BPS saturated spectrum and scattering amplitudes at finite momenta
are not left invariant under S-duality transformations; {\it cf.}\
\HMS.  By relating S-duality transformations in scale invariant $N=2$
field theories to global symmetries in asymptotically free theories, I
will show that these S-dualities are, in fact, {\it exact}
equivalences of their quantum field theories.

The basic idea is to regard the marginal couplings of the scale
invariant theory as lowest components of vector multiplets---complex
scalar ``Higgs'' fields---in an enlarged theory.  Then the coupling
space $\cal M$ of the scale invariant theory is realized as a
submanifold of the Coulomb branch $\cal C$ of the enlarged theory.
Any S-duality identifications of different points of $\cal M$ are
interpreted as equivalences on $\cal C$.  By choosing the enlarged
theory appropriately, these equivalences on $\cal C$ can be made
manifest as (spontaneously broken) global symmetries.

Some elementary properties of the enlarged theory can be deduced
immediately.  The couplings of the enlarged theory must not get strong
in the ultraviolet, for then we would have no control over the Coulomb
branch for large Higgs vevs since it would depend on whatever physics
regulates the theory at short distances.  Furthermore, if the enlarged
theory is itself scale invariant, then the couplings of the smaller
scale invariant theory of interest will typically depend on the
couplings (as well as the Higgs vevs) of the enlarged theory.  One can
find in this way relations among S-duality groups for different
theories \refs{\APS,\AS}, but not the kind of derivation of S-duality
that we are looking for.  Thus we should look for asymptotically free
enlarged theories for which we have exact information about the
Coulomb branch for all values of the Higgs vevs.

In preparation for making the argument outlined above more concrete, I
first briefly describe the (conjectured) S-duality groups of the
theories I will use as examples, and then review the exact solutions
for the Coulomb branches of these models.  However, it should be clear
that this line of argument can be applied to a much wider set of
models than the examples I treat below.  In particular, it would be
interesting to extend it to the series with $Sp(2r)$ gauge group, one
antisymmetric and four fundamental hypermultiplets \ASYT, as well as
to the rich set of recently solved models with product gauge groups
\WiMthNii.  Also, although one cannot derive the S-duality group of
$N=4$ theories in this way from enlarged theories with simple gauge
groups, it may be possible to do so using enlarged theories with
product gauge groups.  Note, in this connection, that there is no
requirement that the enlarged theory have as much supersymmetry as the
scale invariant theory; in particular, one can flow from an $N=2$
supersymmetric theory to an $N=4$ one in the infrared.

\medskip\noindent{\it Some $N=2$ S-dualities.}

In the $N=2$ theories discussed here it is convenient to define the
coupling as $\tau = {\vartheta \over \pi} + i {8 \pi \over g^2}$,
differing by a factor of two from the usual definition.  The $N=2$
theory with $SU(2) \simeq Sp(2)$ gauge group and four fundamental
``quark'' hypermultiplets has S-duality group\foot{I will only discuss
the S-duality action on marginal couplings and not on masses or other
operators, and so will ignore the distinction between $SL(2,\bZ)$ and
$PSL(2,\bZ)$ in what follows.} $SL(2,\bZ)$ \SWii.  The classical
coupling space ${\cal M}_{cl} = \{\hbox{Im}\tau > 0 \}$ is identified
under the transformations $T: \tau \rightarrow \tau +1$ and $S: \tau
\rightarrow -1/\tau$, giving the quantum coupling space ${\cal M} =
{\cal M}_{cl} / SL(2,\bZ)$.  The $SL(2,\bZ)$ duality group can be
presented abstractly as the group with two generators $S$ and $T$
satisfying the relations $S^2 = (ST)^3 = 1$.  The generators $S$ and
$ST$ do not act freely on ${\cal M}_{cl}$, but have fixed points which
are $\bZ_2$ and $\bZ_3$ orbifold points of $\cal M$, if we assign
${\cal M}_{cl}$ a flat metric.  The relations satisfied by the
generators encode the holonomy in $\cal M$ around these orbifold
points.

Since we are free to make coordinate changes on ${\cal M}_{cl}$,
though the topology of $\cal M$ certainly has an invariant meaning, it
may not be clear that the holonomies around points in ${\cal M}$ are
physically meaningful.  In particular, to define holonomies, one needs
a connection.  The existence of a natural connection on coupling space
\JO\ implies that these holonomies are physical \ArPol.  Thus the
S-duality group, considered as an abstract infinite discrete group is
physically meaningful.

The scale invariant $N=2$ theories with a single $SU$, $SO$, or $Sp$
gauge group and quarks in the fundamental (defining) representation
all have low-energy effective theories that are invariant under
identifications of $\tau$ under a discrete group isomorphic to
$\Gamma_0(2) \subset SL(2,\bZ)$ \refs{\APS,\AS}.  $\Gamma_0(2)$ is the
subset of $SL(2,\bZ)$ matrices with even lower off-diagonal entry, or
equivalently, is the subgroup of $SL(2,\bZ)$ generated by $T$ and
$U=ST^2ST$.  The only relation satisfied by these generators is
$U^2=1$, thus characterizing $\Gamma_0(2)$ more abstractly as the
group freely generated by $T$ and $U$ subject to the one relation
$U^2=1$.  Quotienting the classical coupling space by the S-duality
group, we expect a single $\bZ_2$ orbifold point in the quantum
coupling space at the fixed point of the $U$ transformation; see fig.\
1.  An important feature of this duality group is that it does not
remove all the ultra-strong coupling points (the cusp at $\tau=0$ in
the figure) from the quantum coupling space.  No weakly coupled
description of the ultra-strong coupling points is known for the
higher rank theories.
\fig{Fundamental domains in the $\tau$ plane of $\Gamma_0(2)$ (thick 
dashed lines) and $\widetilde\Gamma^0(2)$ (thick solid lines).  Weak
coupling is at $\tau=+i\infty$ and the ultra-strong coupling points
are at $\tau=\{0, \pm1\}$.  The domains are ``folded'' along the
Im$\tau$ axis and opposite edges are identified, giving rise to
$\bZ_2$ orbifold points at $\tau=\{i, \half(1{\pm}i) \}$.}{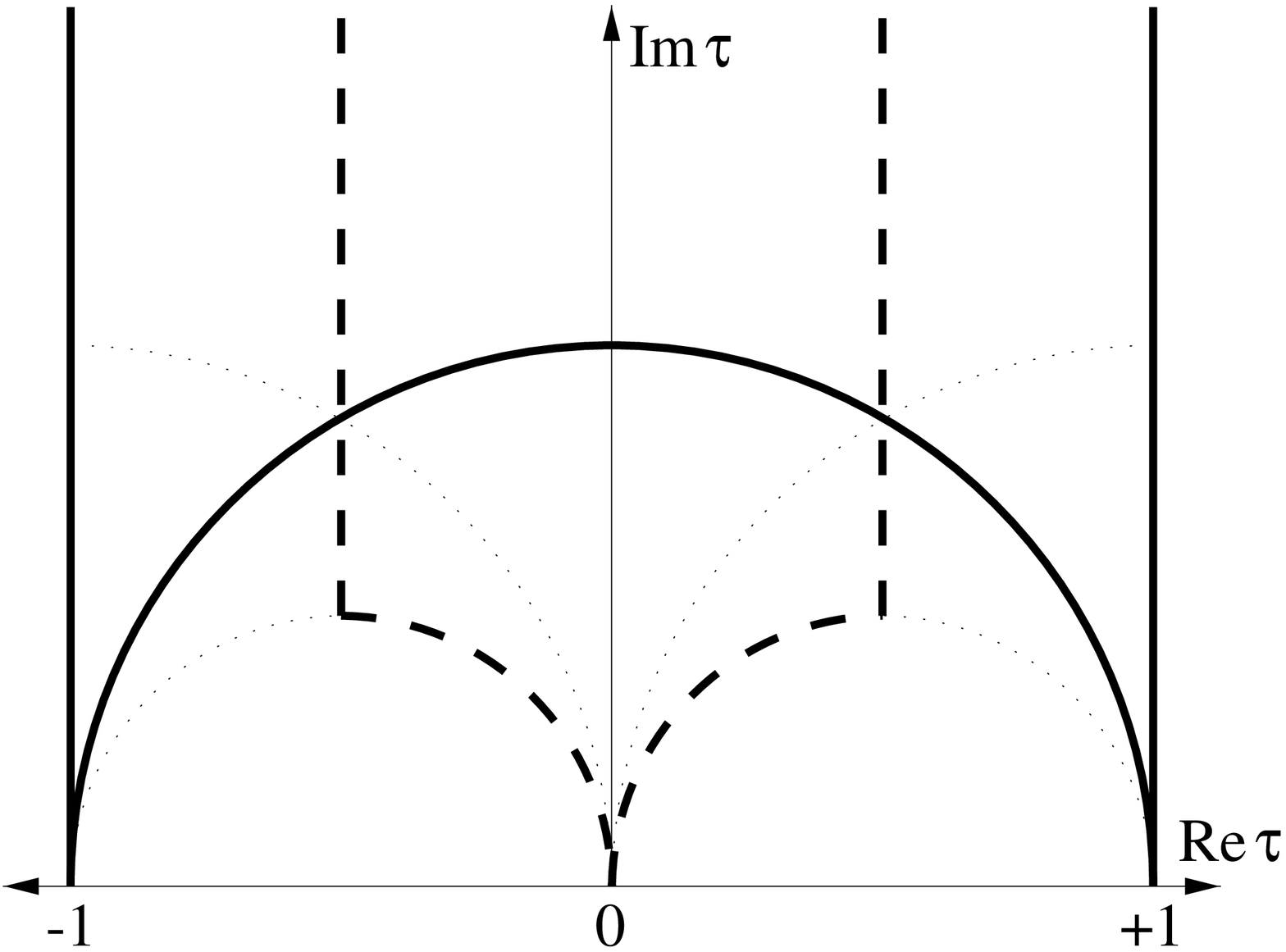}
{7cm}

In fact, in the form of the solution for the effective action on the
Coulomb branch that I use below, only the $Sp(2n)$ series has its
S-duality group realized in precisely this way.  The $SU(n)$ and
$SO(n)$ series realize the duality group instead as the subgroup of
$SL(2,\bZ)$ generated by $S$ and $V=T^2$ which I will denote by
$\widetilde\Gamma^0(2)$.  (It is related to $\Gamma^0(2)$, the subset
of $SL(2,\bZ)$ matrices with even upper off-diagonal entry by a
similarity transformation: $\widetilde\Gamma^0(2) = T \cdot
\Gamma^0(2) \cdot T^{-1}$.)  Its fundamental domain in the $\tau$
plane is also shown in fig.\ 1.  Abstractly it is the freely generated
group on $V$ and $S$ with the one relation $S^2=1$, and so is
isomorphic to $\Gamma_0(2)$.  Nevertheless, the fact that the
S-duality group is realized in these two different ways will be
important below in deriving the full $SL(2,\bZ)$ duality of the $SU(2)
\simeq Sp(2)$ theory.

\medskip\noindent{\it Some exact solutions for Coulomb branches.}

I refer to the complex adjoint scalar component $\phi$ of vector
multiplets as Higgs fields and the two complex scalar components $q$,
$\widetilde q$ of hypermultiplets as quarks.  The Coulomb branch $\cal
C$ is the manifold of degenerate vacua with zero quark vevs and Higgs
vev in the complexified Cartan subalgebra of the gauge group, breaking
$G \rightarrow U(1)^r$ where $r=\hbox{rank}(G)$.  Locally ${\cal C}
\simeq \bC^r$, and I denote its coordinates by $\phi_a$, $a=1, \ldots,
r$.  Note that the point $\phi_a = 0$ will be the scale invariant
vacuum, but that moving away from it along the Coulomb branch to
points with $\phi_a \neq 0$ spontaneously breaks the scale invariance
of the underlying theory and generically flows to the infrared free
$U(1)^r$ Maxwell theory.

The couplings $\tau_{ab}$ of this low-energy effective $U(1)^r$ theory
can receive quantum corrections.  The low-energy $U(1)^r$ theory has
an $Sp(2r,\bZ)$ electric-magnetic duality, implying \SWi\ that the
effective couplings $\tau_{ab}$ are a section of an $Sp(2r,\bZ)$
bundle on $\cal C$.  It turns out that (in every known example) this
$Sp(2r,\bZ)$ ``periodicity'' of $\tau_{ab}$ can be geometrically
encoded by taking $\tau_{ab}$ to be the complex structure (period
matrix) of a genus-$r$ Riemann surface $\Sigma$.  For the models
I will discuss $\Sigma$ is hyperelliptic and so can be
presented as a two-sheeted cover of the complex $x$ plane described
by the complex curve $y^2 = P(x)$ where $P(x)$ is a degree $2r+1$ or
$2r+2$ polynomial.

The curve for the scale invariant $Sp(2r)$ theory with $2r+2$ fundamental 
quarks is \AS
\eqn\Spsi{
	xy^2 = \left[ x\prod_{a=1}^r(x-\phi_a^2) + g \prod_{j=1}^{2r+2}
	m_j \right]^2 -g^2 \prod_{j=1}^{2r+2}(x-m_j^2) .}
(Note that the polynomial on the right side is divisible by $x$.)
Here the quark masses $m_j$ transform in the adjoint of the $SO(4r+4)$
flavor group.  The $\phi_a$ can be taken as the eigenvalues of the
Higgs field $\langle\phi\rangle = \hbox{diag} \{\phi_1, \ldots, \phi_r,
-\phi_1, \ldots, -\phi_r\}$.  A basis of gauge invariant combinations
of the $\phi_a$'s are the symmetric polynomials in $\phi_a^2$ up to
degree $2r$.  $g$ is a function of the coupling which must go as
$8e^{i\pi\tau}$ at weak coupling.  In the scale invariant theory
(setting the masses to zero) the coupling parameter space is thus the
$g^2$ plane; weak coupling is at $g^2=0$, ultra-strong coupling
corresponds to the $g^2=-1$ singularity, while the $\bZ_2$ orbifold
point is at $g^2=\infty$, since it is a fixed point of the $g
\rightarrow -g$ symmetry of the scale invariant curve.  This shows how
the low-energy effective action encodes the S-duality by making its
parameter space isomorphic to a fundamental domain of $\Gamma_0(2)$.
This has been the main evidence for the existence of this duality to
date.

The Coulomb branch of the scale invariant $SU(r+1)$ theory with $2r+2$ 
fundamental quarks is described by the curve \APS
\eqn\SUsi{
	y^2 = \prod_{a=1}^{r+1}(x-\phi_a)^2 + (h^2-1)\prod_{j=1}^{2r+2}
 	(x-\mu_j-h\mu) .} 
Here the quark masses $\mu_j$ transform in the adjoint of the
$SU(2r+2)$ flavor group and satisfy the tracelessness constraint
$\sum_j\mu_j=0$, while the singlet mass $\mu$ is charged under the
$U(1)$ flavor group (``baryon number'').  The $\phi_a$ can be taken as
the eigenvalues of the Higgs field $\langle\phi\rangle = \hbox{diag}
\{\phi_1, \ldots, \phi_{r+1}\}$, subject to the constraint $\sum_a 
\phi_a =0$.  The gauge invariant combinations of the $\phi_a$'s are 
all the symmetric polynomials in $\phi_a$ up to degree $r+1$.  $h$ is
a function of the coupling such that $h \sim 1 + 32e^{i\pi\tau}$ at
weak coupling.  In the scale invariant theory (setting the masses to
zero) the coupling parameter space is the $h^2$ plane;
weak coupling is at $h^2=1$, ultra-strong coupling corresponds to the
$h^2=0$ singularity, while the $\bZ_2$ orbifold point is at
$h^2=\infty$, much as in the $Sp(2r)$ case.

The curves for the scale invariant $SO(n)$ theories \AS\ are quite
similar, and encode the S-duality group in the low-energy effective 
action in an identical fashion.  

In order to derive these S-dualities as exact equivalences of the
entire theory, we will need the effective action for related
asymptotically free theories, obtained by starting from the above
solutions and flowing down by turning on bare quark masses.  In
particular the curve for the asymptotically free $Sp(2r)$ theory with
$2r$ massless flavors is found by making $2r$ masses in \Spsi\ vanish
and setting the remaining two masses to $M \rightarrow \infty$ while
keeping $\Lambda^2 = 8e^{i\pi\tau}M^2 \sim gM^2$ finite.  $\Lambda$ is
the strong coupling scale of the asymptotically free theory.  (I have
chosen the finite factor of proportionality in its definition for
later convenience.)  Thus the massless asymptotically free $Sp(2r)$
theory with $2r$ flavors has a Coulomb branch described by the curve
\eqn\SpAF{
	y^2 = x\prod_{a=1}^r(x-\phi_a^2)^2 - \Lambda^4 x^{2r-1}.}

Similarly, the asymptotically free $SU(r+1)$ curve with $2r$ flavors
is found by setting $2r$ adjoint masses to $-M$, the other two to
$rM$, the singlet mass to $M+\mu$, and taking the $M \rightarrow
\infty$ limit while keeping $\Lambda^2 = -64e^{i\pi\tau}M^2 \sim
(h^2-1)M^2$ finite.  The resulting curve is
\eqn\SUAF{
	y^2 = \prod_{a=1}^{r+1}(x-\phi_a)^2 - (r+1)^2\Lambda^2 
	(x-\mu)^{2r}.}
Here I have kept a finite singlet mass $\mu$ in the asymptotically
free theory for later convenience.

\medskip\noindent{\it Deriving the S-duality of the $Sp$, $SO$ and
$SU$ series.}

We now have the necessary background in place to derive the
conjectured S-dualities of the scale invariant models.  I will start
with the $Sp$ series.  The idea is to take the asymptotically free
$Sp(2r)$ theory with $2r$ quarks as the enlarged theory in which a
scale invariant $Sp$ theory will be embedded.

We can flow to a scale invariant $Sp(2r-2)$ theory by tuning the Higgs
vevs $\phi_a$ so as to break $Sp(2r) \rightarrow Sp(2r-2) \times
SU(2)$ while keeping the quarks massless in the $Sp(2r-2)$ factor.
This is achieved classically by letting one of the $\phi_a$ be
non-zero, say $\phi_r=M$, while setting the rest to zero.  That this
is also true quantumly can be seen by examining the curve \SpAF, which
should factorize into a piece corresponding to the singularity at the
origin of the Coulomb branch of the $Sp(2r-2)$ factor ({\it i.e.}\ the
scale invariant theory we are interested in) and a non-singular piece
corresponding to the $SU(2)$ Yang-Mills factor.  The scale invariant
singularity at the origin of the Coulomb branch for $Sp(2r-2)$ is
simply $y^2 = x^{2r-1}$ as can be seen by setting all the Higgs vevs
and quark masses to zero in \Spsi.  Eq.\ \SpAF\ indeed factorizes in
this way if we set $\phi_r=M$ and the rest to zero:
\eqn\Spfac{
	y^2 = x^{2r-1}\left[(x-M^2)^2 - \Lambda^4 \right].}

The one dimensionless parameter at our disposal in this tuning, namely
$M^2/\Lambda^2$, must be identified with the one dimensionless
parameter (the coupling) of the scale invariant $Sp(2r-2)$ theory.
(One can make this identification more explicit by leaving in finite
masses and Higgs vevs in the asymptotically free $Sp(2r)$ theory, and
taking $\Lambda$ and $\phi_r=M$ to infinity keeping their ratio fixed.
In the limit one directly obtains the scale invariant $Sp$ curve
\Spsi\ with $g = \Lambda^2/M^2$.)

We have thus derived the quantum coupling space of the scale invariant
$Sp$ theories as the complex $M^2$ plane.  Note that $M^2 \rightarrow
\infty$ corresponds to weak coupling since then the asymptotically free
enlarged theory breaks in the ultraviolet.  There are also two special
points at $M^2 = \pm \Lambda^2$ where the term in square brackets in
\Spfac\ has a zero.  These we identify with new (ultra-)strong coupling
physics as they imply the scale invariant $Sp(2r-2)$ theory includes
additional degrees of freedom at those couplings.  In summary, the
quantum coupling space of the scale invariant $Sp$ theories seems to
have one weak coupling and two ultra-strong coupling points.  This
does not match with the conjectured $\Gamma_0(2)$ S-duality of these
theories which we saw above is equivalent to having a quantum coupling
space with one weak coupling, one ultra-strong coupling and one $\bZ_2$
orbifold point.

However, the asymptotically free theory has a global (non-anomalous)
$\bZ_2$ R-symmetry which acts on the $M^2$-plane as $M^2 \rightarrow
-M^2$ ({\it i.e.}\ it is spontaneously broken away from the origin of
the Coulomb branch).  The $M^2$ plane and hence the quantum coupling
space of the scale invariant theory is to be divided by the action of
this symmetry, which then gives one weak coupling point at infinity,
one ultra-strong coupling point at $M^4= \Lambda^4$, and a $\bZ_2$
orbifold point at the origin.  We have thus derived the S-duality of
the $Sp$ series.

(In the case of non-zero bare masses, the $\bZ_2$ global symmetry used
above is the one described in \SWii\ for $SU(2)$.  It is a subgroup of
the non-anomalous $\bZ_8$ found by an appropriate combination of a
global $U(1)_R$, the $\bZ_2$ parity of the $O(4r)$ flavor group, and a
global $SU(2)_R$ transformation.  It acts on the Higgs and quark
fields by $\phi_a \rightarrow e^{i\pi/2} \phi_a$, $q_1 \rightarrow
e^{-i\pi/4} \widetilde q_1$, $\widetilde q_1 \rightarrow e^{-i\pi/4}
q_1$, $q_j \rightarrow e^{-i\pi/4} q_j$, and $\widetilde q_j
\rightarrow e^{-i\pi/4} \widetilde q_j$.)

Very similar manipulations can be applied to the $SO$ series of
models.  By tuning Higgs vevs in an asymptotically free $SO(n)$ theory
with $n-4$ massless flavors to give the scale invariant $SO(n-2)$
theory, one derives its conjectured S-duality group in essentially the
same way.

The $SU$ series is slightly different because one must also tune a
non-zero quark mass in the asymptotically free $SU(r+1)$ theory with
$2r$ quarks to find the scale invariant $SU(r)$ theory.  Classically
we achieve this breaking by tuning the Higgs vevs so that $\phi_a = M$
for $1\le a \le r$ and $\phi_{r+1} = -rM$ (by the tracelessness
condition).  In addition we must tune the singlet mass to $\mu= M$ to
keep the $2r$ quarks massless.  These tunings are also valid quantumly
since upon applying them to the asymptotically free $SU$ curve
\SUAF, and shifting $x \rightarrow x+M$, it factorzes as
\eqn\SUfac{
	y^2 =x^{2r}\left[ (x+(r+1)M)^2 - (r+1)^2\Lambda^2 \right],}
and we recognize the $x^{2r}$ factor as the singularity of the scale
invariant $SU(r)$ theory with $2r$ massless quarks.  We must identify
the dimensionless parameter $M/\Lambda$ which we have tuned with the
coupling of the scale invariant theory; one can explicitly check from
the form of the curves that $\Lambda^2/M^2 = 1-h^2$.  {}From the
degenerations of \SUfac\ the quantum coupling space has a
weak coupling point at $M=\infty$ and two ultra-strong coupling points
at $M = \pm \Lambda$.  Furthermore there is a non-anomalous $\bZ_2 \in
U(1)_R$ which acts on the Higgs fields as $\phi \rightarrow -\phi$
(and when appropriately combined with a global flavor rotation takes
$\mu \rightarrow -\mu$), so that the $M$ plane is identified under $M
\rightarrow -M$, giving a single ultra-strong coupling point and a
$\bZ_2$ orbifold point at $M=0$, thus deriving the content of the
conjectured S-duality for the $SU$ series.

\medskip\noindent{\it Deriving the $SL(2,\bZ)$ duality of the $SU(2)\simeq
Sp(2)$ theory.}

Though both the $Sp$ and $SU$ series include the $SU(2) \simeq Sp(2)$
theory with four fundamental quarks, their S-dualities derived above
do not include the full $SL(2,\bZ)$ duality argued to hold in this
theory \SWii.  For example, upon specializing the $SU$ scale invariant
curve \SUsi\ to the $SU(2)$ case, one finds that the curve is
invariant under the following redefinition of the coupling $h$ \APS:
\eqn\hredef{
	U: h \rightarrow {1-h\over 1+3h} }
(with an accompanying $SL(2,\bC)$ transformation on the $x$
coordinate).  When combined with the $\widetilde\Gamma^0(2)$ duality
derived above, this gives rise to the full $SL(2,\bZ)$
duality---namely, a quantum coupling space with a weak coupling point,
a $\bZ_2$ orbifold point, a $\bZ_3$ orbifold point, and {\it no}
ultra-strong coupling point.  (This is easiest to see in the $h$ plane
where there are weak coupling points at $h=\pm1$ and an ultra-strong
point at $h=0$.  The $\widetilde\Gamma^0(2)$ duality is then generated
by $V: h \rightarrow -h$.  The combined group generated by $U$ and $V$
is isomorphic to $SL(2,\bZ)$.)  It is clear, however, that the
nonlinear transformation \hredef\ cannot be realized as a global
symmetry in the enlarged asymptotically free $SU(3)$ theory.

Nevertheless, we can deduce this nonlinear equivalence from a global
symmetry in another way.  Consider instead the $SU(2)$ theory as part
of the $Sp$ series.  We derived the $\Gamma_0(2)$ duality of this
model above, realizing the redefinition
\eqn\gredef{
	U': g \rightarrow -g }
of the coupling $g$ of the $Sp(2)$ theory as a global symmetry of the
asymptotically free $Sp(4)$ theory.  Now, the equivalence of the scale
invariant $SU(2)$ and $Sp(2)$ curves \SUsi\ and \Spsi\ implies that
their coupling parameters must be related by
\eqn\ghrel{
	h = {g+1 \over 3g-1} .}
(See \AS\ for details, especially section 3.4.)  Combining \gredef\
with \ghrel\ gives \hredef, and thus derives the full $SL(2,\bZ)$
duality of the $SU(2) \simeq Sp(2)$ scale invariant theory.  Thus, by
enlarging the $SU(2)$ theory in two different ways, different
generators of its S-duality group are realized as global symmetries
which together generate the full group.  This can be understood
pictorially by noting that the intersection of the fundamental 
domains of $\Gamma_0(2)$ and $\widetilde\Gamma^0(2)$ in fig.\ 1 is a 
fundamental domain of $SL(2,\bZ)$.

It is a pleasure to thank A. Buchel and A. Shapere for many helpful
discussions.  This work was supported by NSF grant PHY-9513717.

\listrefs
\end